\documentclass[11pt]{article}

\usepackage{graphicx} 
\usepackage{subcaption}
\usepackage{float}
\usepackage[ruled,lined,linesnumbered,commentsnumbered]{algorithm2e}  
\usepackage[most]{tcolorbox}
\usepackage{lipsum}
\usepackage[letterpaper, top=1in, bottom=1in, left=1in, right=1in]{geometry}
\usepackage{minitoc}
\usepackage{tablefootnote}
\usepackage{authblk}
\usepackage{xcolor}
\usepackage{minitoc}
\usepackage[utf8]{inputenc} 
\usepackage[T1]{fontenc}    
\usepackage{hyperref}       
\usepackage{url}            
\usepackage{booktabs}       
\usepackage{amsfonts}       
\usepackage{nicefrac}       
\usepackage{microtype}      
\usepackage{xcolor}  
\usepackage{wrapfig} 
\newcommand{\hl}{\textcolor{red}}
\usepackage[final]{acl}
\usepackage{titletoc}
\usepackage{times}
\usepackage{latexsym}

\usepackage[T1]{fontenc}

\usepackage[utf8]{inputenc}

\usepackage{microtype}

\usepackage{hyperref}  
\usepackage{xcolor}   

\hypersetup{
    colorlinks=true,  
    linkcolor=red,       
    urlcolor= orange
}

\usepackage{inconsolata}

\usepackage{graphicx}

\title{Language Models Should be Used to Surface\\the Unwritten Code of Science and Society}

\author{
  Honglin Bao\textsuperscript{†},
  Siyang Wu\textsuperscript{†},
  Jiwoong Choi\textsuperscript{*},
  Yingrong Mao\textsuperscript{*},
  James A. Evans
\\
\\
  University of Chicago
\\
  \small{† HB and SW co-led the project. * JC and YM contributed equally.}
\\
}

\begin{document}

\maketitle
\begin{abstract}
This position paper calls on the research community not only to investigate how human biases are inherited by large language models (LLMs) but also to explore how these biases in LLMs can be leveraged to make society’s "unwritten code" — such as implicit stereotypes and heuristics — visible and accessible for critique. We introduce a conceptual framework through a case study in science: uncovering hidden rules in peer review - the factors that reviewers care about but rarely state explicitly due to normative scientific expectations. The idea of the framework is to push LLMs to speak out their heuristics through generating self-consistent hypotheses - why one paper appeared stronger in reviewer scoring - among paired papers submitted to 46 academic conferences, while iteratively searching deeper hypotheses from remaining pairs where existing hypotheses cannot explain. We observed that LLMs’ normative priors about the internal characteristics of good science extracted from their self-talk, e.g., theoretical rigor, were systematically updated toward posteriors that emphasize storytelling about external connections, such as how the work is positioned and connected within and across literatures. Human reviewers tend to explicitly reward aspects that moderately align with LLMs’ normative priors (correlation = 0.49) but avoid articulating contextualization and storytelling posteriors in their review comments (correlation = –0.14), despite giving implicit reward to them with positive scores. These patterns are robust across different models and out-of-sample judgments. We discuss the broad applicability of our proposed framework, leveraging LLMs as diagnostic tools to amplify and surface the tacit codes underlying human society, enabling public discussion of revealed values and more precisely targeted responsible AI.
\end{abstract}

\section{Introduction}

Large language models (LLMs), extensively trained on vast corpora of human-generated text and visuals, inevitably absorb and reflect the patterns of human thought embedded within these datasets. Researchers have explored how AI can be used to judge human activities such as scientific research \citep{liang2024can}, misinformation oversight \citep{chen2024humans}, and perceptions of physical appearance \citep{conwell2025perceptual}. With strong predictive capacities, LLMs can replicate shifts in public opinion and human beliefs on contentious political issues, including same-sex marriage \citep{kim2023ai} and climate change \citep{lee2024can}. These models often reveal biases that closely align with directly measured human biases and even amplify them across various social dimensions \citep{ziems2024can}. While this position paper moves beyond merely acknowledging LLM biases; it calls on leveraging them as diagnostic tools to uncover and examine the implicit, unwritten societal codes that govern human interaction.

\begin{figure}
    \centering
    \includegraphics[width=0.45\textwidth]{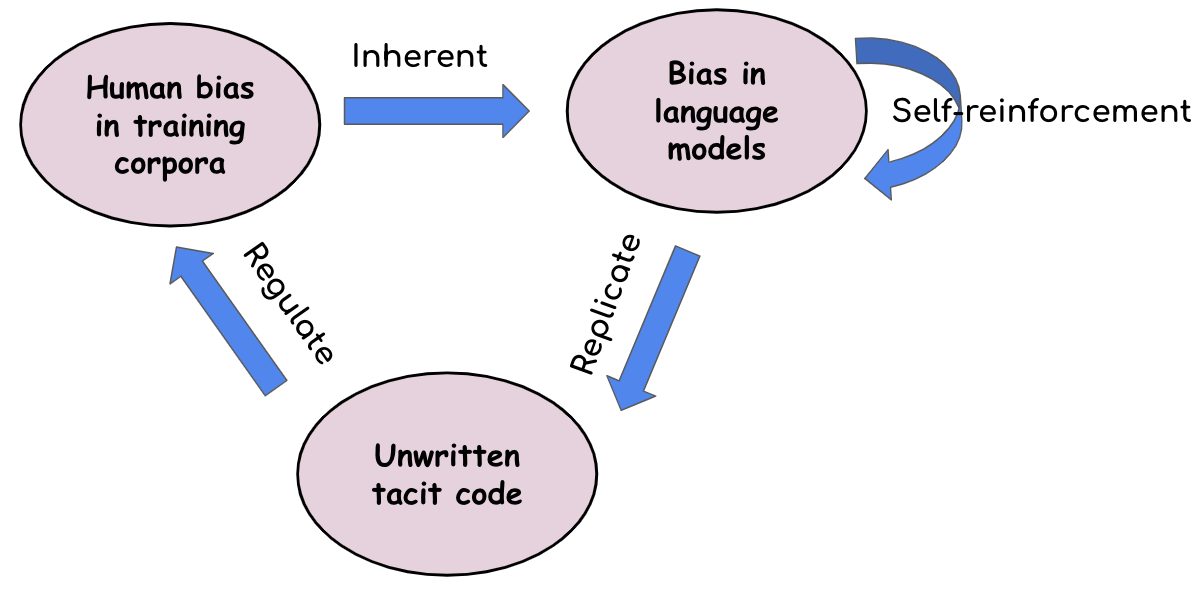} 
    \caption{We argue and present a conceptual framework to self-reinforce biases to amplify unwritten codes until the model articulates and replicates them explicitly.}
    \label{fig:overview1}
\end{figure}

The ``unwritten code" -- comprising tacit norms, implicit stereotypes, and subtle conventions that invisibly influence societal dynamics -- represents a potent force that, if left unexamined, unsummarized, and unaddressed, can perpetuate systemic inequities and injustices. 
Leveraging the inherent biases embedded within LLMs can shed light on otherwise elusive patterns that sustain societal inequality. One straightforward path to mitigating human bias involves summarizing and clearly articulating the biases humans hold, subsequently enabling evaluation, regulation and systemic change. Nevertheless, these biases frequently exist "unwritten" and thus resist formalization and systematic analysis. Even though some methods can expose bias in LLMs when models appear to avoid expressing it, as by analyzing associated words \citep{bai2025explicitly}, there remains a deeper layer of unwritten code that contemporary simple techniques cannot recover. These deeper structures activate during the model’s reasoning process, serving as heuristics that streamline its efficiency \footnote{We have more discussions in Section \ref{diss} and Appendix \ref{appendix:related_work}.}. An algorithmic framework capable of extracting and explicitly articulating them through self-reinforcement and amplification of the rationale within LLMs, which are inherently imbued with human biases and heuristics, could illuminate the content and character of bias within models and human cultures. Thus, this position paper calls for an intentional shift in perspective: \textbf{We call on researchers not only to investigate how human biases are inherited by LLMs but also to explore how these biases can be leveraged to make the unwritten codes visible}, as shown in Figure~\ref{fig:overview1}. To this end, we present a conceptual framework that enables this process by pushing LLMs to speak out their heuristics through searching their deeper self-consistent hypotheses that can explain their decision-making. By prompting models to think out of the box and "exaggerate" hidden rules that humans might be unable or unwilling to acknowledge openly, we can more clearly illustrate how such biases and heuristics operate and which groups they potentially disadvantage.

\section{A Case Study: The Aesthetic of Peer Review}

What makes a scientific paper "good"? While scientific communities often emphasize norms such as rigor, the actual process of evaluating quality is far more entangled with social dynamics, implicit biases, and tacit expectations \citep{bao2024simulation}. Peer review, the central gatekeeping mechanism of science, is deeply shaped by these social dynamics. Despite its appearance of objectivity, peer review can be biased toward well-known institutions, established authors, or fashionable topics \citep{lee2013bias}. Reviewers may assign high scores to work that aligns with unspoken disciplinary tastes or collective values \cite{sulik2025differences}, while "lazily" masking these preferences in the language of standardized evaluation criteria \citep{purkayastha2025lazyreview}. As a result, much of what constitutes the tacit code of a field -- its evaluative culture -- remains hidden beneath the surface, shaping scientific recognition in nontransparent ways that remain hard to alter. 

\textbf{\textit{Data}}: We collected data from the OpenReview API and Paper Copilot \footnote{https://papercopilot.com/}, a website that aggregates peer review scores (mostly) for computer science conference submissions. Our final dataset includes metadata, peer review scores, and review comments for 26,731 papers submitted to 46 conferences including computer science, physics, medicine, and social sciences. We use peer review scores as a proxy for the paper's perceived quality, as shown in Appendix \ref{sec:review}, due to the consistency and high confidence of judgments among reviewers.

To represent each paper’s content, we follow the approach of \citep{zhang2024massw}, constructing an extended abstract that integrates additional structured elements — contextual background, key ideas, methodological and theoretical details, experiments and results, and mentioned impact. We generate these extended abstracts using prompts used in \citep{zhang2024massw} with GPT-4o. This representation generation has been empirically shown to achieve high quality across evaluation metrics and to closely match human annotations. It also demonstrates strong alignment with the original text and a low rate of hallucination \citep{amar2023openasp, zhang2024massw}, see Appendix \ref{extended} for fidelity experiments. We deliberately avoid feeding full texts directly to LLMs for two main reasons. First, context windows remain limited, preventing broad coverage that would enable more generalizable hypotheses. Current approaches, e.g., \citep{movva2025sparse}, employ reduction methods such as sparse autoencoders to compress long inputs into activated neurons for hypothesis generation, but these are typically applied only to straightforward cases (e.g., political attitudes) where activations can be reduced to obvious dimensions (e.g., gender and race). Second, prior work has shown that LLMs often struggle with long, unstructured inputs, leading to degraded performance and increased hallucination rates mid-text \citep{liu2024lost, yen2024helmet}. When presented with lengthy texts, attention may diffuse across irrelevant sections, impairing reasoning over core content. To address these issues, we adopt a curated and condensed input format, ensuring that the model attends to the most relevant information. Dataset and paper representations are detailed in Appendix~\ref{appendix:ap2}. Most importantly, robustness checks in the Appendix \ref{extended} show that LLM judgments on full papers and extended abstracts yield highly consistent results (consistency = 0.89).

\section{The Framework: Prior, Posterior, and Hypothesis Generation}

\label{frame_text}

We adopt a hypothesis generation framework -- a topic increasingly explored in the automation of scientific discovery using AI \citep{radensky2024scideator, liu2024literature, batista2024words, liu2025hypobench} -- to update LLMs' stereotyped priors about good science into posteriors that reflect how human evaluators and LLMs actually judge. 

\begin{figure*}[htbp]
    \centering
    \includegraphics[width=0.8\textwidth]{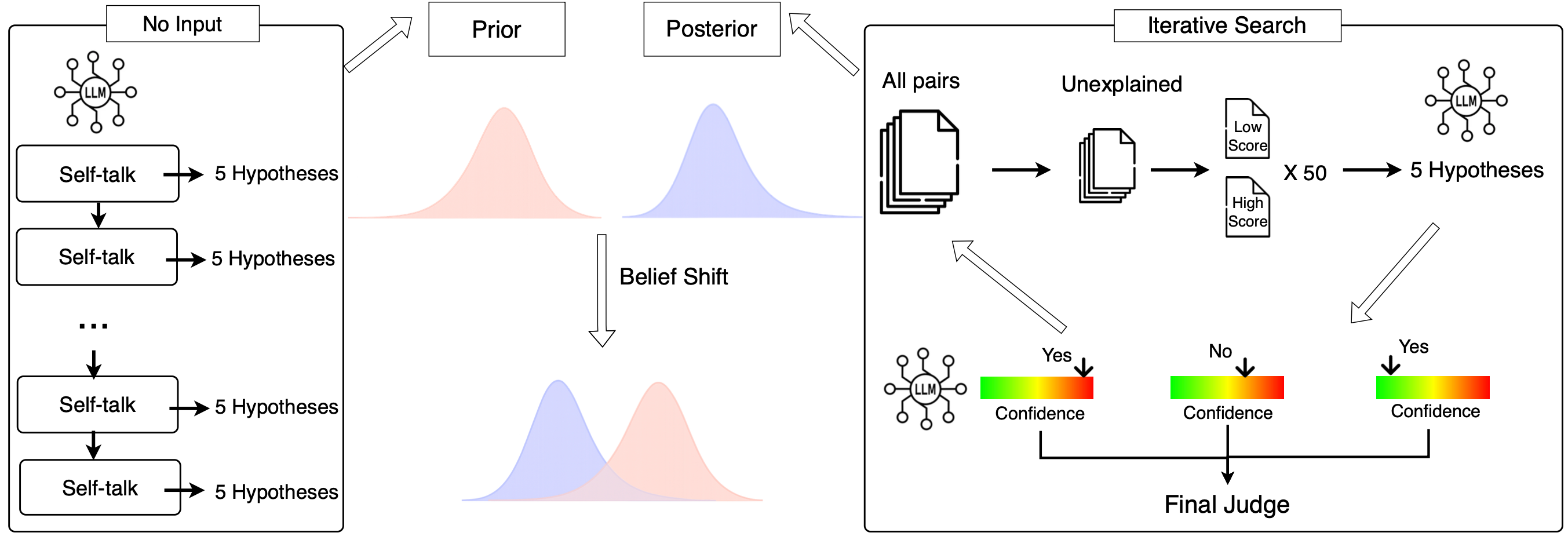} 
    \caption{Updating priors to posteriors by hypothesis search.}
    \label{fig:overall2}
\end{figure*}

\textbf{Hypothesis Generation}: We begin by constructing a pairwise dataset. For each conference, we identify all pairs of papers where one paper has received a significantly higher review score than the other. To reduce noise due to randomness, we define a "significant" score gap as one that exceeds one standard deviation within the score distribution of that conference. From this pool, we randomly sample 50 paper pairs (due to the limited contextual window) and feed them into an LLM. The LLM is tasked with reading, understanding, and comparing the papers' extended abstracts, and then generating five hypotheses to explain why the better scoring paper may appear stronger than the other.

\textbf{Self-consistency Refinement}: We apply the five generated hypotheses to all pairs in the dataset. For each pair and hypothesis, we ask the LLM to determine the degree to which the hypothesis explains the observed quality difference (human review scores) using the following prompt:
"Analyze the content from both papers to determine whether Paper 2 is superior in the aspects identified in the hypothesis (i.e., if the comparison of content supports the hypothesis)." This type of self-confirmation explicitly reveals the inherent biases of LLMs during the judgment process. The LLM outputs two things: (1) a binary label (0 or 1) indicating whether the hypothesis explains the score difference, and (2) a confidence score (0–10) for its judgment. We use 3-fold voting and compute the final label using a confidence-weighted scheme: if the sum of confidence scores for the label "1" exceeds that of label "0", the final label is 1; otherwise, it is 0. This confidence-weighted approach has been shown to improve self-consistency and accelerate convergence to the stable answer of LLMs across a variety of QA tasks \citep{taubenfeld2025confidence}. Beyond improving reliability, confidence scores across multiple judgments provide a fine-grained signal that distinguishes borderline cases from clear-cut ones - an especially valuable property in our setting, where evaluating internal characteristics is inherently difficult \citep{bao2024simulation} and confidence levels can help identify when the model is genuinely uncertain \citep{taubenfeld2025confidence}. We also tested majority voting, and the resulting labels are highly correlated with the confidence-weighted labels (correlation > 0.98). LLMs may exhibit position bias, meaning that when serving as judges, they tend to favor the candidate appearing in a specific position (e.g., the first) \citep{shi2024judging}. We adopt standard strategies to mitigate this bias. First, we construct paper pairs with substantial quality differences; prior research has shown that position bias is the most pronounced when the gap is small \citep{li2024split, shi2024judging}. Second, we randomize the positions of Paper 1 and Paper 2 across voting while maintaining logically consistent prompts and voting aggregation. Consistency of judgments across randomized positions achieves 77\% and we have more discussions in Appendix \ref{appendix:position}.

\textbf{Iterative Search}: The initial set of hypotheses may not account for all observed differences. For cases where none of the hypotheses yield a confident explanation (i.e., the confidence-weighted label is 0 for all hypotheses), we designate them as "unexplained". In each round, we sample 50 such unexplained pairs and prompt the LLM to generate five new hypotheses that are distinct from the existing ones. This iterative process continues until the proportion of unexplained cases drops below 5\%. Note that each newly generated hypothesis is evaluated across all paper pairs, not just the subset of unexplained instances from which it was derived. Although new hypotheses are generated from a specific subset, they may still provide valid explanations for other pairs previously explained by different hypotheses. In our experiments, we find that 4 rounds of search, yielding 20 hypotheses, cover 97\% of pairs under judge model agreements. This suggests that scientific judgments can often be explained by a compact set of generalizable hypotheses, reflecting recurring evaluative patterns across diverse cases \footnote{In Section \ref{diss} and the results below, we further decompose these generalized hypotheses into priors (which humans tend to explicitly emphasize) and posteriors (humans’ implicit judgments).}.

\textbf{Prior and Posterior}: From a Bayesian perspective, querying an LLM with a general question ``What do you think makes one paper appear stronger than another?'' without providing any specific data input, elicits the model’s \textbf{prior} $P(\text{hypothesis})$. This prior reflects the LLM’s internal distribution over explanatory hypotheses about what makes a scientific paper strong, independent of evidence. To extract this prior, we prompted the model to generate hypotheses in the absence of data. Specifically, we conducted 4 rounds per simulation, with 5 distinct hypotheses generated in each round. Repeating this simulation 250 times yielded a total of 5{,}000 \text{prior hypotheses}, following the same procedure of iteratively searching hypotheses. When the model is shown actual data, such as comparisons between paper pairs, it updates its beliefs, resulting in a \textbf{posterior} distribution $P(\text{hypothesis} \mid \text{data})$ over hypotheses, which captures how likely the model believes a given hypothesis explains the observed data and how confidently it makes the judgment. Operationally, we estimate this posterior by computing, for each hypothesis, the proportion of paper pairs it can successfully explain when the model is exposed to actual inputs.

We are particularly interested in belief change during this process, i.e., how LLM's prior about good science transforms into a robust posterior, after judging the actual pairwise dataset. For each generated hypothesis, when the LLM judges the pairwise dataset, we retain the proportion of cases it can explain (one pair can be explained by multiple hypotheses). We also match each hypothesis to priors (prompting: do you think they are talking about the same idea and same hypothesis?) to obtain their appearance frequency in the prior. We also conducted cross-validation experiments for the reliability of LLM labels of "match" using a heuristic dictionary-based method (locating some keywords of posterior hypotheses in priors to identify "match"), and these two methods exhibit a relatively high degree of correlation of 0.70, $p$=0.001 (see Appendix \ref{appendix:dict} for details). The frequency of each posterior hypothesis in the prior distribution is measured by counting how often a similar hypothesis appears within each window of 20 sequential prior hypotheses across 250 simulations. We use 20-hypothesis windows to align with the posterior generation setup, where LLMs are prompted to produce 4 rounds of 5 unique hypotheses. This ensures a fair comparison between the frequency of hypotheses in the prior and posterior distributions.

Overall experiments are formalized in Appendix \ref{overall_al}. Across experiments, we used OpenAI o3-mini for hypotheses (naively generate priors and iteratively generate posteriors in hypothesis search). We used OpenAI GPT-4o to match prior and posterior hypotheses, i.e., using a good reasoning model to "think" and a high-capacity aligned model to give a stable and reliable judgment. The overall framework of this process can be seen in Figure~\ref{fig:overall2}. All prompts used in the paper can be viewed in Appendix~\ref{appendix:prompts}. Generating high-quality summaries for roughly 27K papers, pairing them under our criteria to create more than 14 million pairs, and having an advanced LLM to read, hypothesize, and evaluate each pair multiple rounds would lead to prohibitive computational and monetary costs. For demonstration purposes in our case study, we randomly selected 5,000 pairs for the main experiments. To strengthen the robustness of our conclusions, we further evaluated an additional 5,000 pairs through two robustness checks: (1) whether they produce similar hypothesis patterns, and (2) whether the generated set of hypotheses based on one dataset can be applied to out-of-sample data. Detailed discussions in Appendix \ref{outofsample} show that our conclusions are both generalizable and robust. Furthermore, in Appendix \ref{other_model} we replicate our experiments across a range of popular closed- and open-source LLMs, all of which yield qualitatively similar patterns.

\section{Results}

We highlight the key results in the main paper. To comprehensively test generalizability and robustness, we designed 13 sections of experiments in the appendix. Across models, out-of-sample data, human annotation, and alternative representations of papers, we find that the major conclusions consistently hold.

\subsection{Prior and posterior manifest distinct generation logics}

\begin{figure}[htbp]
    \centering
    \begin{subfigure}[t]{0.21\textwidth}
        \centering
        \includegraphics[width=\linewidth]{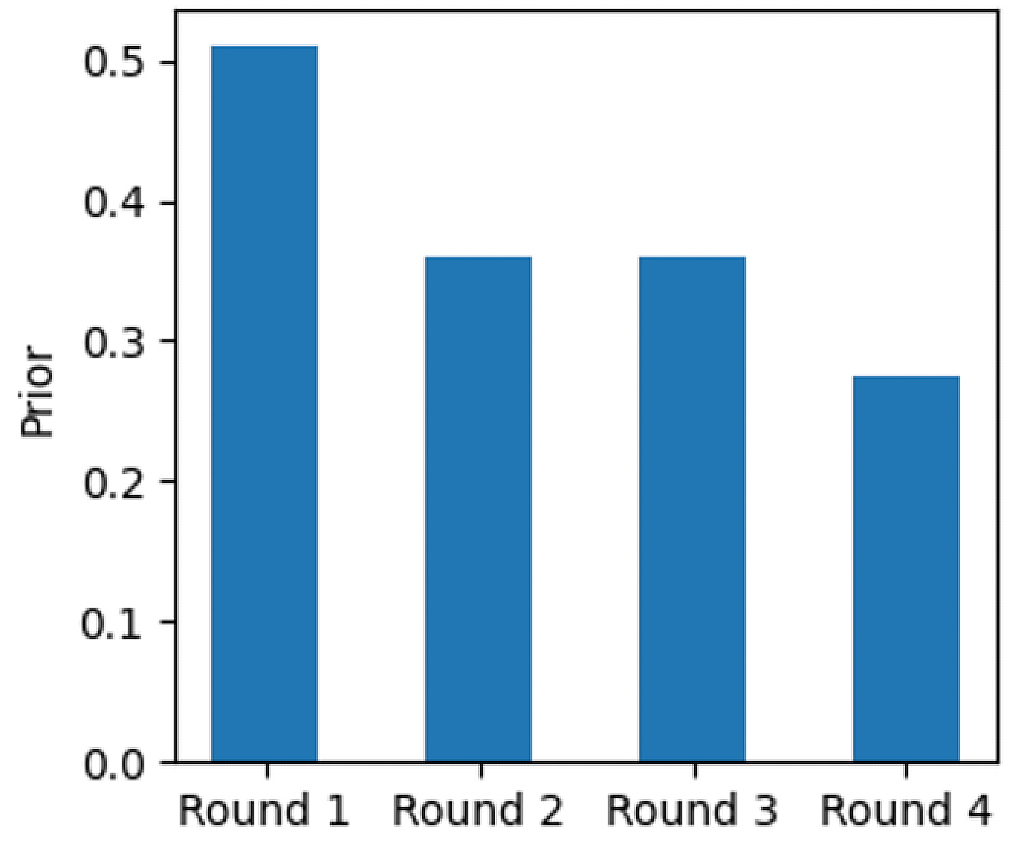}
        \caption{The appearance of the generated hypothesis in the priors}
        \label{fig:understand}
    \end{subfigure}
    \hspace{0.02\textwidth}
    \begin{subfigure}[t]{0.2\textwidth}
        \centering
        \includegraphics[width=\linewidth]{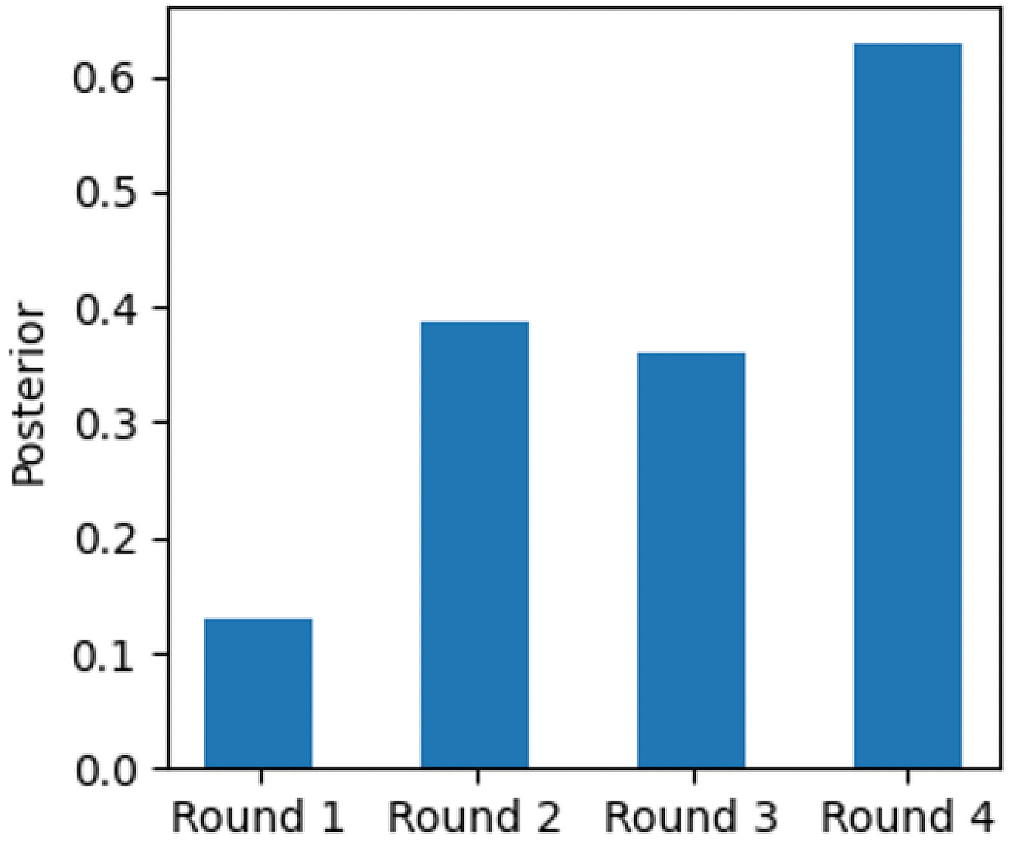}
        \caption{The proportion of explained pairs by generated hypothesis (posterior) across rounds}
        \label{fig:distinguish1}
    \end{subfigure}
  
    \begin{subfigure}[t]{0.2\textwidth}
        \centering
        \includegraphics[width=\linewidth]{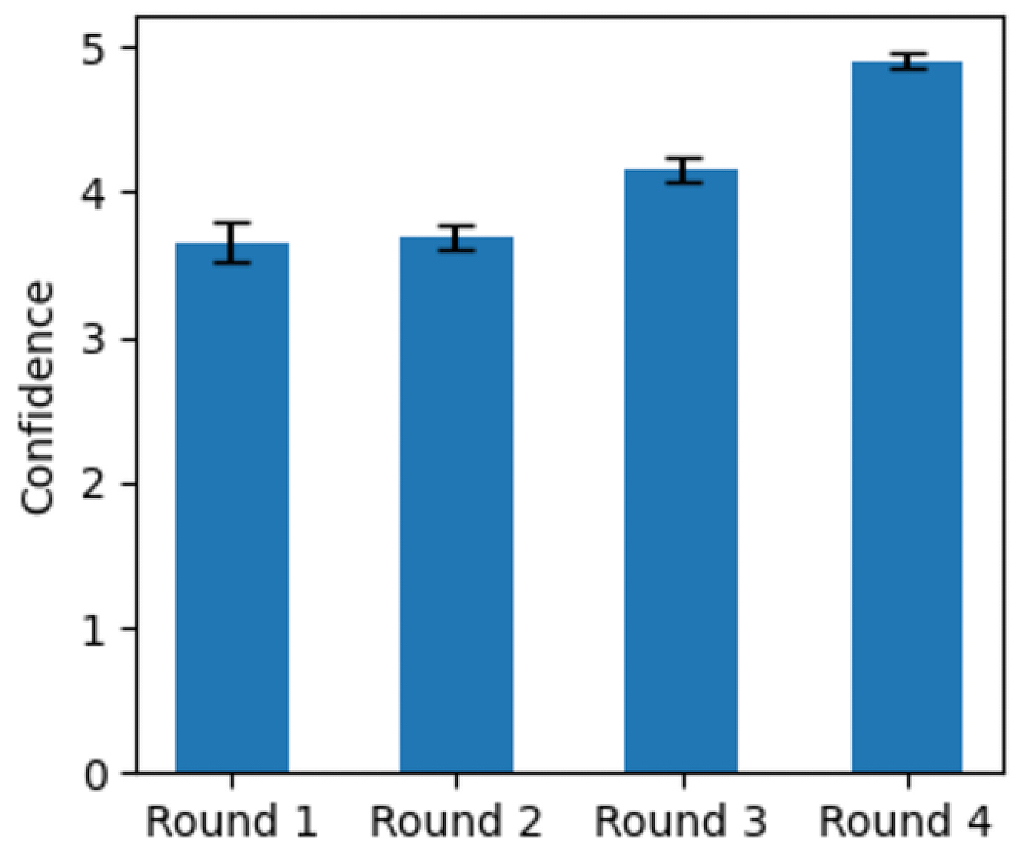}
        \caption{LLM's increased confidence when making the judgment}
        \label{fig:distinguish2}
    \end{subfigure}
    \hspace{0.02\textwidth}
    \begin{subfigure}[t]{0.2\textwidth}
        \centering
        \includegraphics[width=\linewidth]{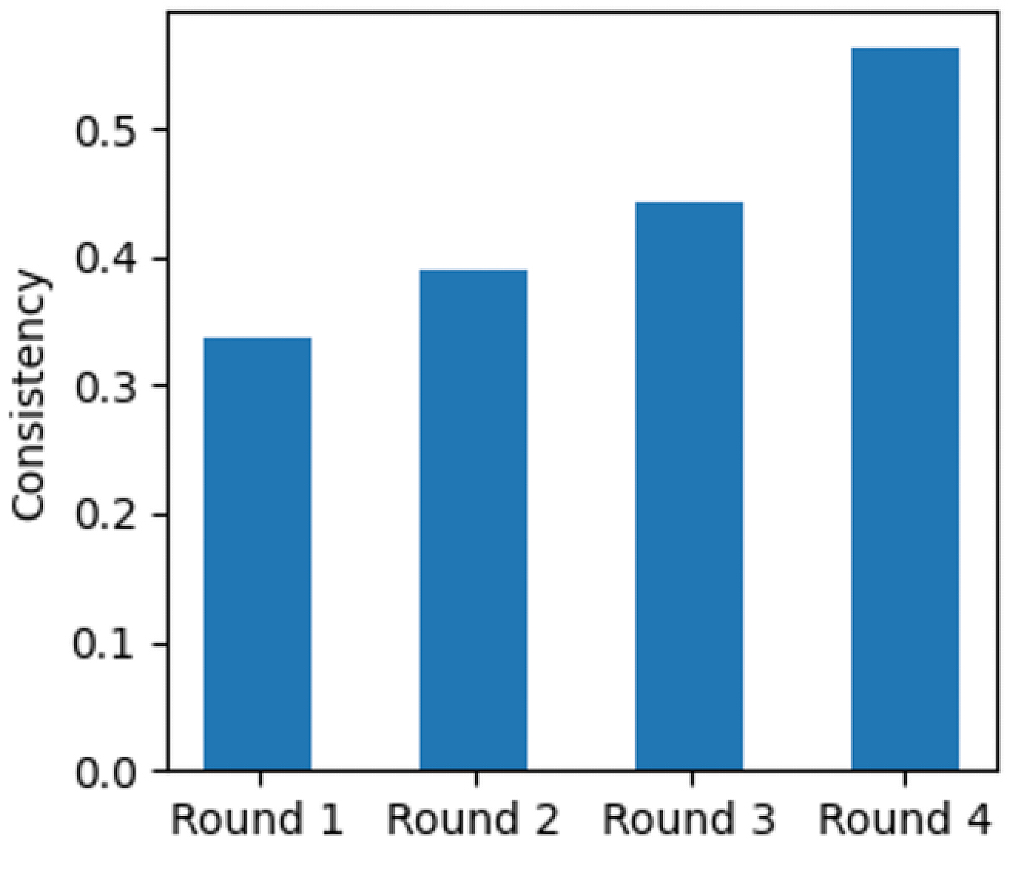}
        \caption{LLM's increased consistency when making the judgment}
        \label{fig:distinguish3}
    \end{subfigure}
    
    \caption{Prior and posterior manifest distinct generation logics.}
    \label{fig:combined2}
\end{figure}

We first examine how hypotheses are generated. Generation occurs over four rounds, each producing five hypotheses. As shown in Figure~\ref{fig:combined2}, the frequency with which hypotheses appear in the model’s prior belief gradually decreases across rounds (represented by the mean of each round). This trend makes intuitive sense: the LLM initially generates hypotheses that are most "normative" for the standard of good science, such as theoretical rigor. These are the kinds of stereotyped considerations that naturally come to mind when researchers judge scientific work. As the rounds progress, we prompt the model to think beyond common ideas, encouraging the generation of less frequently seen hypotheses that can still explain peer review decisions in previously unexplained cases. Interestingly, these later hypotheses exhibit stronger explanatory power across the dataset. That is, although they are generated from cases that were initially hard to explain, they nevertheless prove applicable to cases already covered by earlier, more normative hypotheses. Moreover, the LLM shows greater confidence — quantified as the difference between the total confidence of votes matching the final weighted label and that of votes for the alternative label across three rounds — and improved consistency, measured by the proportion of pairs in which LLM casts the same vote across all rounds. Initially, the judging LLM draws on its priors but lacks confidence in assessing factors such as theoretical rigor. The model shows greater confidence in later rounds when evaluating the newly generated hypotheses, however, which are largely centered on presentation, contextualization, and storytelling (we will return to these hypotheses in the next section). The full list of generated hypotheses is presented in Table~\ref{table:hypos} in the appendix. These hypotheses are also distinct, with the pairwise cosine similarity in posterior coverage (1–0 sequences) being 0.43 $\pm$ 0.02, indicating that they cover relatively different dimensions of the scientific quality signal. 

\subsection{Human judgments and LLMs share priors, but not posteriors}

\begin{figure*}
    \centering
    \includegraphics[width=0.7\textwidth]{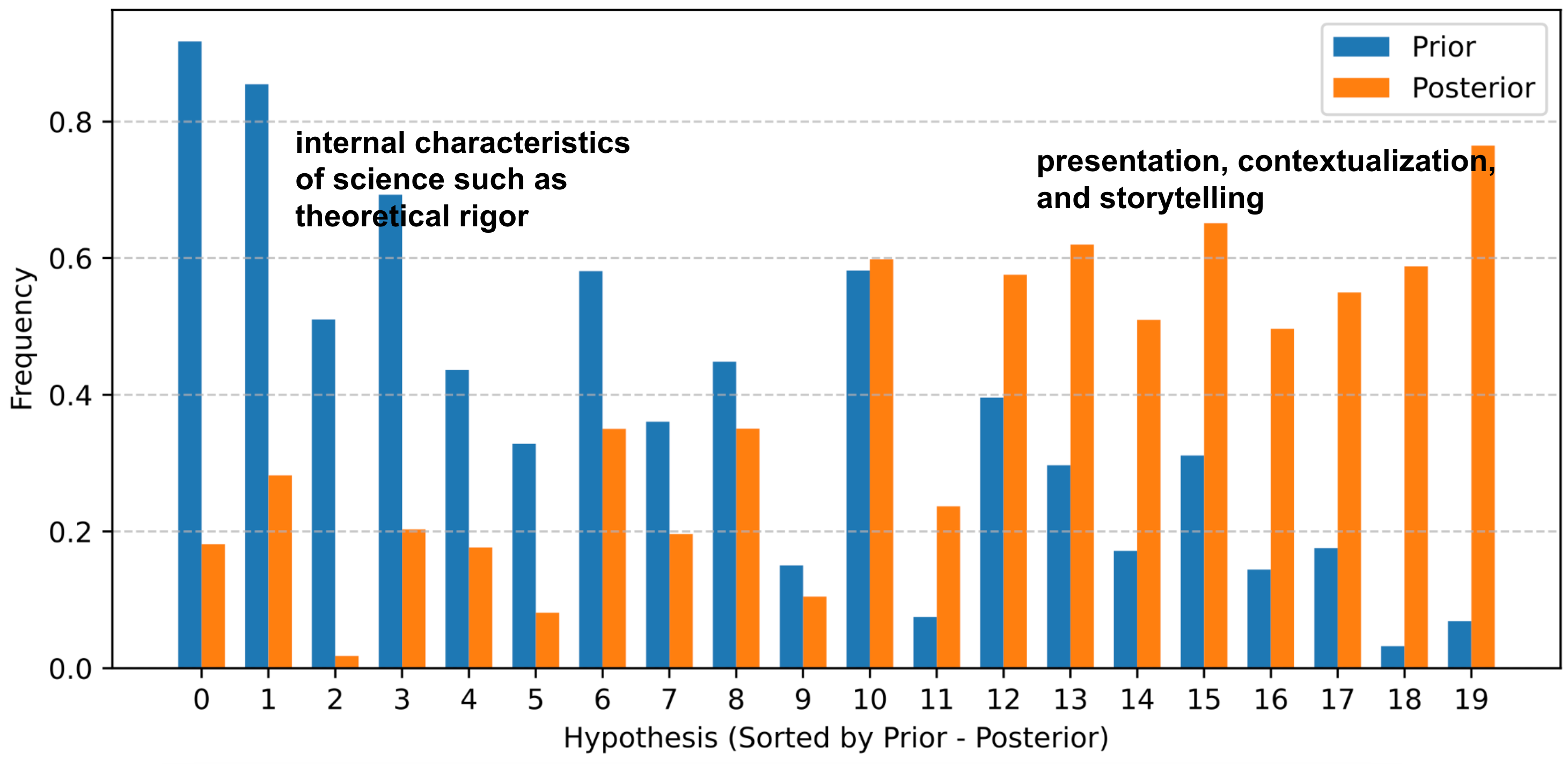}
    \caption{Prior versus Posterior. The index of each hypothesis represents its rank in the change of the prior frequency minus the posterior frequency (from high to low).}
    \label{fig:overview}
\end{figure*}

We observe notable shifts in the frequency of certain hypotheses from prior to posterior, as shown in Figure \ref{fig:overview}. Some hypotheses remain relatively stable in frequency and their explanatory power for good science (middle). In contrast, others manifest substantial change -- some decrease markedly in frequency (left), while others increase significantly (right) from the prior distribution (see Table \ref{table:hypos}). This shift reflects a transition, after being exposed to the pairwise data, from emphasizing normative scientific values -- such as theoretical or methodological rigor - to prioritizing more practical, communicative, and contextualization aspects, including storytelling quality and interdisciplinary relevance. When treated as a peer reviewer, the LLM initially aligns with conventional scientific ideals. In practice, however, it leans more heavily on accessible, narrative-driven criteria that appear more explanatory in distinguishing lower from higher rated papers (Figure \ref{fig:combined2} panels c and d). This suggests a divergence between explicit and implicit criteria of evaluation in science, a phenomenon widely appreciated although rarely measured in social studies of science \cite{latour1987science}.

We collected human review comments and annotated them using in-context learning prompts with GPT-4o, leveraging its strong performance in social computing annotation tasks, often surpassing human experts and crowd-workers \citep{tornberg2023chatgpt, alizadeh2023open, haq2025llms}. For each hypothesis, the annotation model was provided with three illustrative examples of reviewers not-mentioning (label 0) specific aspects in a hypothesis, along with the reviewers’ attitudes toward those aspects (positive: 1 or negative: -1, they are considered as "mentioning"). For robustness, we conducted two experiments: 1) the dictionary method produces high consistency with LLM's "mention" label with correlation of 0.78, $p=0.000$, detailed in Appendix \ref{appendix:dict}. 2) Three master’s students specializing in social computing independently conducted the human annotations, achieving a high level of consistency with the LLM-generated labels (see Appendix \ref{annotation} for details). We find human comments align with LLM’s priors in terms of the frequency of aspects they address, meaning that they articulate a similar attention distribution as LLM's prior. In contrast, human reviewers rarely articulate their updated judgments like the LLM does. That is, if we assume the LLM mirrors human reviewing behavior, both begin from similar priors. However, while the LLM transparently updates its evaluation (posterior), human reviewers often refrain from articulating such shifts -- likely due to normative expectations in academic reviewing -- despite favoring unarticulated values with higher scores (we have more discussions and empirical evidence in Section \ref{diss}). 

Supporting this, the top five hypotheses in Table \ref{table:hypos} with the largest prior-to-posterior gains (e.g., storytelling, orange texts) account for 61\% of the LLM’s posterior attention, but only 15\% of its prior and 21\% of mentions in human feedback. Conversely, the top five hypotheses with the largest prior-to-posterior losses (e.g., internal rigor, blue texts) account for 17\% of the LLM’s posteriors, but as much as 68\% of its prior and 50\% of human mentions. LLMs, as expected, initially even exaggerate stereotyped evaluation norms — overweighting "prior" and underweighting "posterior" standards — more sharply than humans do. Humans indeed explicitly reward internal qualities, but their stated reward patterns explain very little of the actual score distribution ($R^2$ = 0.06, details about the regression analysis can be seen in Appendix \ref{appendix:regression}), suggesting the presence of implicit rules guiding the evaluation process. Human reviewers tend to explicitly comment on aspects that moderately align with the LLM’s normative priors (mention frequency and prior frequency correlation = 0.49) but are much less likely to articulate storytelling and contextualization posteriors in their written feedback (mention frequency and posterior coverage correlation = –0.14). All these suggest a possibility that while these storytelling aspects gain importance during evaluation — filling in the missing explanatory power $R^2$ of regressions — they remain underreported in human reviews, perhaps because they deviate from traditional scholarly norms, despite their relevance for establishing significance.




\section{Discussions}
\label{diss}

It is important to emphasize that our goal is not to advocate for the automation of judgment tasks by LLMs. Rather, our primary goal is to \textbf{present a position through a conceptual framework} — namely, to exploit the biases and preference rationale embedded in LLMs by human discursive culture, then self-amplifying, reproducing, and explicitly identifying who might be harmed by these biases. Much of the existing research has focused on more explicit, easily identifiable forms of biases associated with social dimensions such as gender and race \citep{ntoutsi2020bias}. Researchers have developed methods to bypass the barriers when the model refuses to disclose its biases, such as using associated words \citep{bai2025explicitly}. However, they constitute the “prior” itself. These biases and heuristics still manifest superficially and can be uncovered through direct or indirect cues. In contrast, what we focus on in this paper is the "posterior" and how it diverges from the prior. Surprisingly, the results show that even when an LLM's prior (i.e., its stereotypes) is "good", its posterior can still be "bad" - let alone when its prior is already "bad". When acting as reviewers, LLMs appear to draw on two distinct systems to explain differences in quality signals. One is more normative and less confident in the evaluation process. The other is more narrative-driven, context-sensitive, and ultimately more confident and consistent. When judgment can be achieved through both systems, LLMs are more likely to apply the second system across the board as heuristics because they more consistently explain quality differentials. Humans, like LLMs, may also favor contextual criteria in evaluation, but they articulate this shift in terms of normative standards. 

We make our position dialectical with alternative perspectives and policy suggestions:

\textbf{How to articulate the unwritten code in other high-stakes evaluative settings?} In hiring, LLMs can be asked to rank resumes: while initial judgments may rely on normative criteria such as education, experience, and skills, prompting models to articulate deeper rationales can surface less explicit but influential factors like perceived cultural fit—for instance, inferring misalignment when a candidate’s prior employers differ culturally from the hiring firm \citep{rivera2012hiring} and being rejected silently. Similarly, in college admissions, encouraging LLMs to explain their evaluations of personal statements enables systematic analysis of implicit reasoning; prior work shows that essays from certain ethnic or linguistic backgrounds exhibit stylistic differences that may be inadvertently penalized \citep{lee2025poor, alvero2021essay, alvero2024multilingualism}. Examining LLM scoring rationales in these contexts can thus help identify how implicit signals contribute to bias in hiring, admissions, and criminal justice.

\textbf{To what extent does an LLM’s shift from prior to posterior actually reflect human reasoning?} Humans may never explicitly confess to using narrative or contextual heuristics in reviewing and other social dimensions, yet we have fairly strong evidence that they do. Prior survey-based studies show that reviewers in the social sciences place substantially greater emphasis on “interpretive challenges that question a study’s framing” than on methodological rigor \citep{teplitskiy2016frame, strang2015revising}. We push LLMs to articulate such emphasis and generalize this insight to conference papers in technical fields, where - owing to normative expectations around methodological rigor and SOTA performance - reviewers are less inclined to make such concerns explicit \footnote{https://hackingsemantics.xyz/2020/reviewing-models/} but use "lazy" words as a mask precisely in NLP \citep{purkayastha2025lazyreview}. Moreover, if human score variance could be understood as comprising both explicit and implicit components, with LLMs’ prior-to-posterior shift accounting for nearly all variance (as shown in Section \ref{frame_text} "iterative search") while human-stated explicit judgments map only to the prior, then the posterior can reasonably be taken to capture the implicit part.



Another view holds that even if we acknowledge the existence of unwritten codes, \textbf{making them explicit serves little purpose and may even enable people to “game the system.”} But this view neglects the crucial balance between transparency and normativity. The fear that publicizing unwritten codes will be exploited implicitly assumes that it is preferable for such rules to continue operating in the shadows. In reality, the opposite is true: "hidden curriculum" benefits those who already occupy advantaged positions or happen to be in the know \citep{hariharan2019uncovering}. For example, researchers from elite institutions who are familiar with Western academic norms naturally understand how to write and frame a compelling “story” \citep{chen2024geographical} - they have long benefited from tacit rules without needing them to be made explicit. The purpose of transparency is not to provide a cheat sheet, but to allow marginalized groups to understand the hidden barriers they face and thereby push institutions to reflect on and revise those standards. If people do game the system after unwritten codes are revealed, it is not a failure of transparency — it is evidence that those codes themselves need institutional correction. 

\textbf{Can LLMs replace the peer review and other judge processes?} Our goal is not to provide a definitive answer -- although certainly not yet. Nevertheless, our work suggests that insofar as LLMs function as cultural technologies that encode societal norms \citep{blodgett2020language, farrell2025large}, then their "cognitive shift" in peer review reflects analogous unacknowledged shifts in human reviewers. That said, while LLM-generated reviews have been proven useful and cover broader and more specific aspects than human reviews \citep{liang2024can, yuan2022can, thakkar2025llm} -- likely including those humans may be unwilling or unable to disclose -- LLM reviewers still lack constructive feedback essential for critique when judging a standalone paper. Pairing papers for comparative judgment of feature existence and magnitude in some sense addresses this concern, as demonstrated here and in previous experiments \citep{bai2025explicitly}.

\textbf{What should we do at the implementation level?} LLMs cannot be made fully unbiased, but they can be designed to surface, label, and contrast the implicit heuristics they rely on. At an implementation level, this means shifting from single-shot judgments to procedures that explicitly elicit multiple rationales, track prior–posterior shifts, and expose which criteria become dominant only after iterative comparison. Practically, LLMs should be used as diagnostic tools rather than automated judges: systems can require models to generate and evaluate multiple competing hypotheses, retain confidence signals, and highlight systematic gaps between what is explicitly articulated and what is implicitly rewarded. These signals should be logged and presented to human decision-makers, not collapsed into final scores. We should establish evaluation metrics for transparency, not just accuracy. Implementation should emphasize perspective diversity. Using multiple models, prompts, or reviewer roles (e.g., complementary–neutral–proximal perspectives) helps prevent any single normative frame from dominating. Diversity here should be enforced not only at the level of prompts, but at the representational level - through steering, architectural tweaks, or training differences that lead to meaningfully different internal representations and behaviors. Such collective calibration mechanisms allow institutions to recognize, debate, and potentially revise their unwritten codes.

\newpage

\section*{Limitations}

In this paper, we focus on conference papers. Future research could examine more selective, longer, and detailed review comments from journals, although such data are often subject to legal constraints. Conducting an anonymous survey across both technology and social science fields would further strengthen the proposed framework. Our study employs well-normalized and venue-controlled scores; future work could incorporate more nuanced statistical analyses on factors such as institutional prestige.

\newpage

\bibliography{references}    

\newpage

\appendix
{\Huge Appendix}

\section{Reviewer score is a good measure of quality signal}
\label{sec:review}

\begin{figure}[htbp]
    \centering
    \begin{subfigure}[t]{0.28\textwidth}
        \centering
        \includegraphics[width=\linewidth]{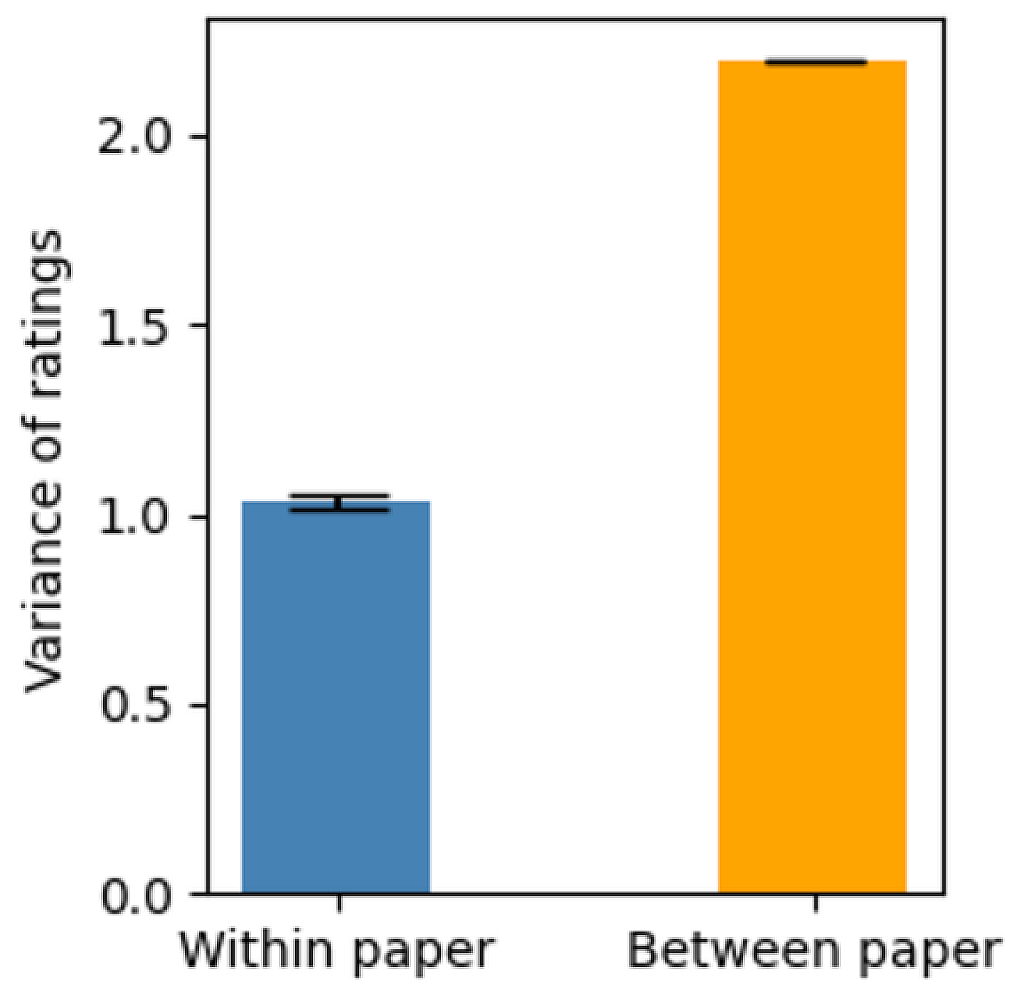}
        \caption{The judgment is consistent}
        \label{fig:perplexity}
    \end{subfigure}
    \hspace{0.05\textwidth}
    \begin{subfigure}[t]{0.3\textwidth}
        \centering
        \includegraphics[width=\linewidth]{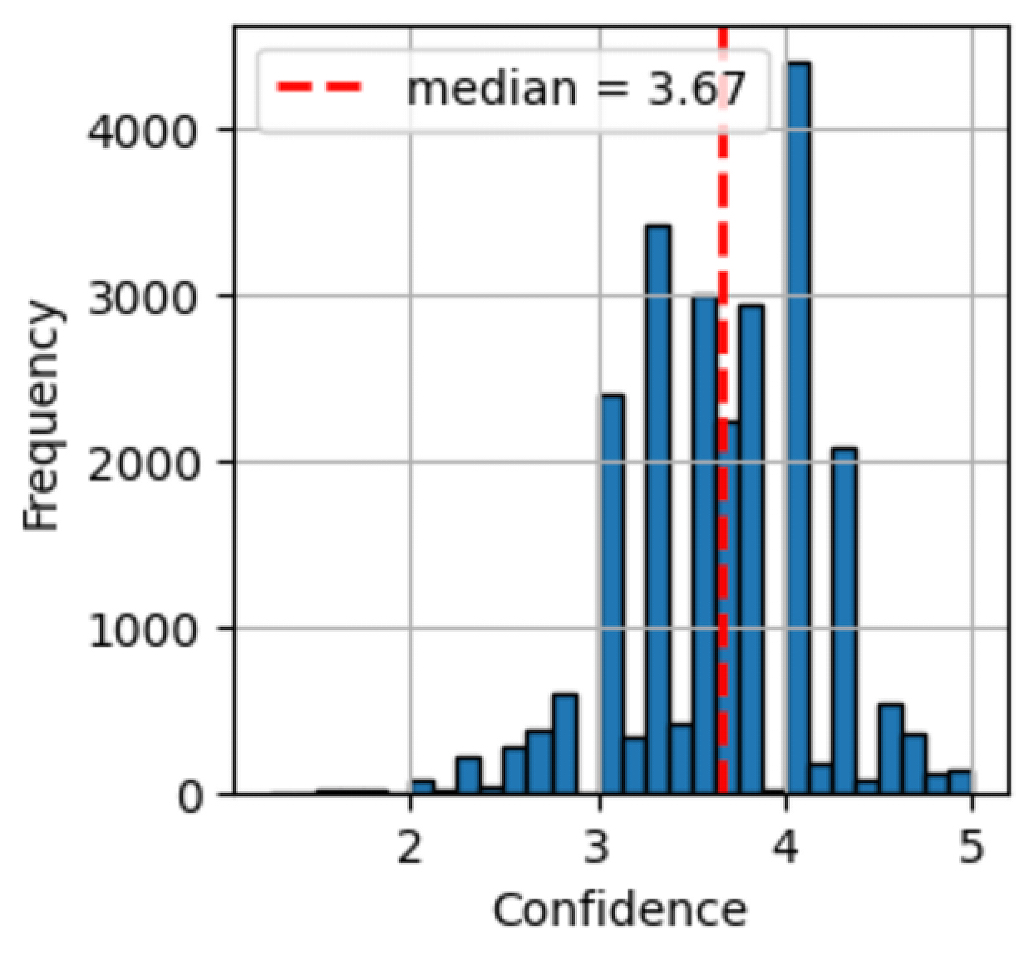}
        \caption{The judgment is confident}
        \label{fig:similarity}
    \end{subfigure}
    \caption{Review score is a good measure for quality.}
    \label{fig:combined}
\end{figure}

Review scores tend to be much more consistent within a single paper than across different papers (see Figure~\ref{fig:combined} panel a), suggesting that multiple reviewers held shared values, enabling consensus in their evaluations. Reviewers reported relatively high confidence in their judgments with a median of 3.67 out of 5 (see Figure~\ref{fig:combined} panel b). Thus, in general the review score is a good measure of quality signal.

\begin{figure}[htbp]
    \centering
    \includegraphics[width=0.38\textwidth]{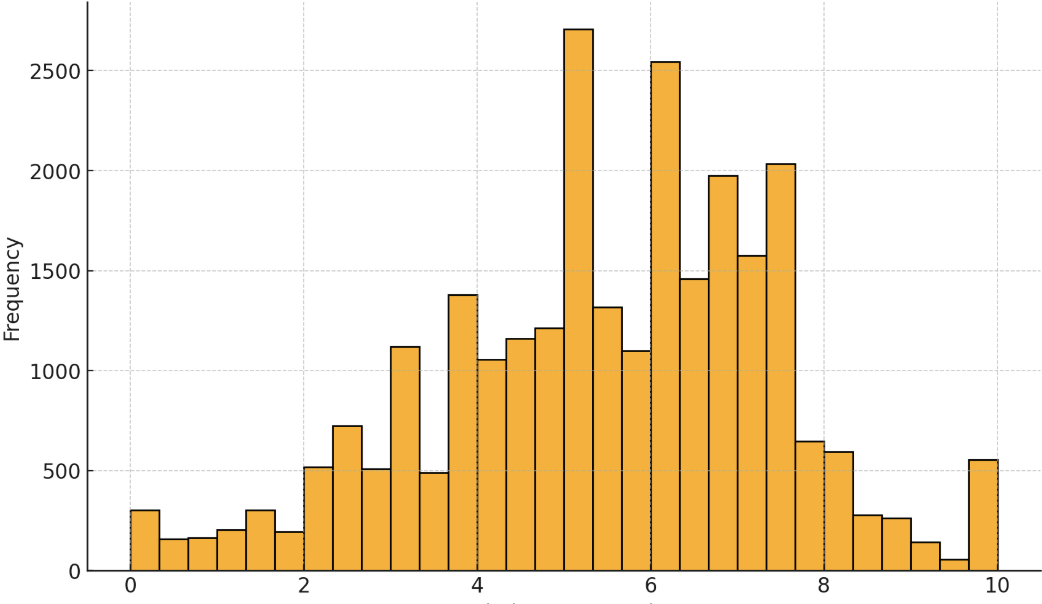} 
    \caption{Review scores are normally distributed.}
    \label{fig:score}
\end{figure}

\section{Human annotation experiments}
\label{annotation}

Here, we complement the labeling methods (LLM labeling (major) and the dictionary method (complementary)) with human annotation. Three master’s students specializing in social computing were involved to label the reviewer comments using exactly the same prompt as GPT-4o (if reviewers’ attitudes toward the aspects in a given hypothesis were positive = 1 or negative = –1, they were considered as “mentioning”, or 0 - not mentioned at all). The annotators are from Asia: two males and one female. They labeled independently, without using AI assistance and without communicating with one another. We find that one person shows high consistency with LLM labels, with an overlap of 0.76. Their majority-voted label across annotators shows an overlap with LLM labels of 0.70. These results show the reliability of LLMs in annotating text data.

\section{Results using other LLMs are consistent}
\label{other_model}

Across the paper, we intentionally adopted models from the same company (OpenAI) because we wanted the entire process to be shaped by a consistent set of cultural biases, making it more straightforward to expose and analyze these biases.

To increase the generalizability of our work, we explored other popular open-sourced or closed models, including Deepseek-chat-v3-0324 \footnote{https://api-docs.deepseek.com/news/news250325}, Mistral 24b instruct \footnote{https://huggingface.co/mistralai/Mistral-Small-24B-Instruct-2501}, and Gemini 2.5 Flash-Lite \footnote{https://docs.cloud.google.com/vertex-ai/generative-ai/docs/models/gemini/2-5-flash-lite. The result of Gemini 2.5 Flash is very similar to that of its lite version.}, to replicate the full set of experiments. Overall, we find our conclusions to be robust. Researchers have observed increasing convergence among LLMs, largely because major model families are trained on similar mixtures of large-scale web corpora, code repositories, and curated texts \citep{wu2025mapping, wolfram2025layers}. If several LLMs from different companies independently articulate the same belief-change dynamics, this convergence itself may be informative for the biases learned from in-the-wild, human-generated corpora. Specifically:


\begin{itemize}
    \item Alignment between priors and human comments, but not posteriors: All three additional models show a relatively consistent set of priors as OpenAI models used in the main paper, and align with human comments about the definition of “good science”, compared to the low alignment for posteriors VS human comments. Human reviewers tend to explicitly reward aspects that moderately align with LLMs’ normative priors: OpenAI correlation = 0.49; Deepseek 0.28; Gemini 0.65; Mistral 0.45. By contrast, human reviewers generally avoid articulating posteriors in their review comments (OpenAI correlation = - 0.14; Deepseek = –0.16; Gemini = -0.20; Mistral = -0.22), but all models do articulate these posteriors.
    \item Shift from prior to posterior: During the transition from prior to posterior, we still observe a systematic shift from rigor-related aspects toward storytelling perspectives. All models demonstrate surprisingly high consistency in using certain heuristics such as “whether the paper tells a good interdisciplinary story”, and across models, they show on average 89\% agreement with Open-AI’s judgment. This hypothesis also predictably gained the highest prior-to-posterior attention gain: interdisciplinary integration (OpenAI top-1; DeepSeek top-1; Gemini top-2; Mistral top-5) and contextualization (DeepSeek top-2; OpenAI top-2; Gemini top-1; Mistral top-1) are both systematically overestimated in the posterior belief shift. By contrast, rigor is consistently underestimated (OpenAI lowest top-1; DeepSeek lowest top-1; Gemini lowest top-2; Mistral lowest top-2).
\end{itemize}



\section{Out-of-sample judgments}
\label{outofsample}

For robustness, we experimented with another set of 5,000 pairs. We find the generated posteriors in the main paper can still explain a substantial proportion (>95\%) of pairs, even though the models had not seen them at all. Moreover, these posterior distributions are highly correlated (0.80) with those on the 5000 pairs in the original paper. We also experimented with using LLMs to generate hypotheses from the new dataset. We found that the generated hypotheses are qualitatively consistent, with only minor variations in order. We briefly noted this in the original paper (see Section \ref{frame_text} iterative search). We conclude that scientific judgments can often be explained by a compact set of generalizable hypotheses—spanning both easy heuristics and harder-to-discern, quality-driven priors—reflecting recurring evaluative patterns across diverse cases.

\section{Fidelity experiments}
\label{extended}

\begin{figure}[htbp]
    \centering
    \includegraphics[width=0.48\textwidth]{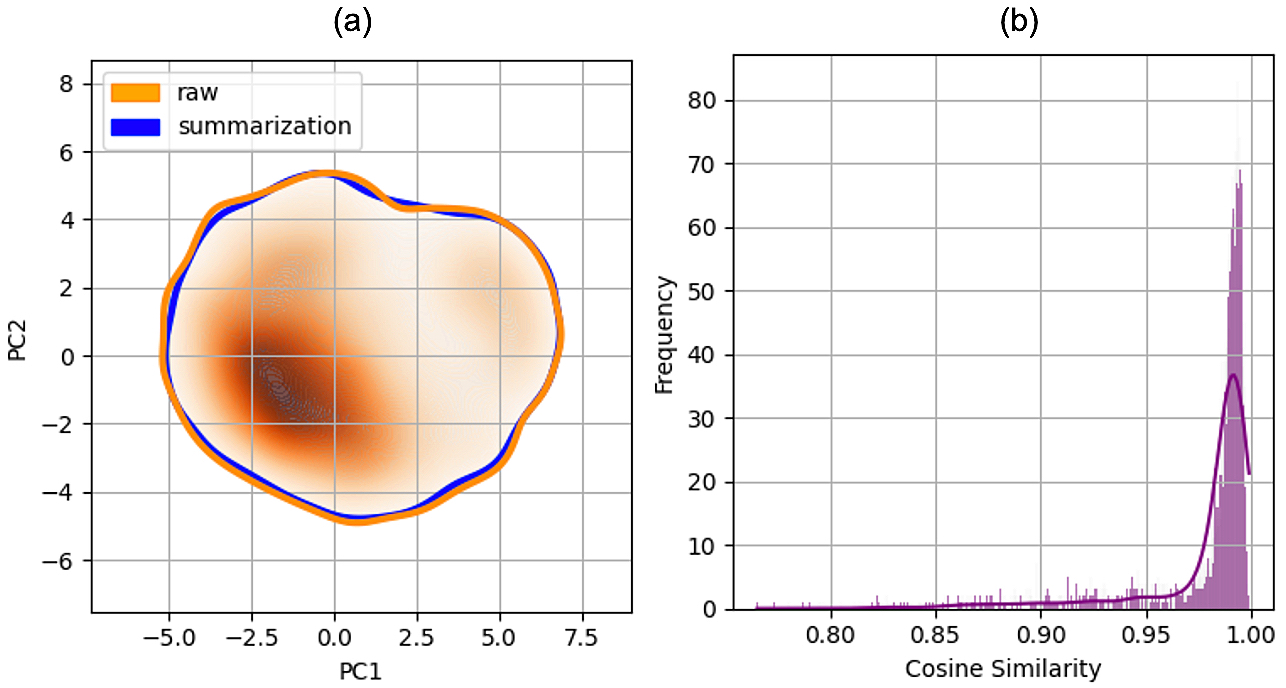} 
    \caption{The summarization exhibits minimal hallucination (panel a - almost overlapping embeddings) and a high degree of fidelity (panel b).}
    \label{fig:sim}
\end{figure}

Summarization using methods from \citep{zhang2024massw} demonstrates high fidelity compared to the raw text. In Figure \ref{fig:sim}, panel (a), we use the embedding model from \cite{singh2022scirepeval}, specifically pretrained on scientific papers' citation relations using contrastive learning. We embed both the summarizations and their corresponding raw texts in each section. To visualize and compare the distribution of these embedding areas, we reduce their dimensionality to two. We then estimate the boundary of each region using a 5\% Gaussian kernel density estimate, drawing contours that enclose the densest 95\% of data points in each embedding set. If the summarizations contain significant hallucinations, we would expect their embeddings to cover meaningfully different regions than those of the raw texts. However, in panel (a), we find that the distributions largely overlap, suggesting strong semantic alignment. In panel (b), we compute the cosine similarity between the two (raw, summarized) embeddings and find that most values are above 0.95.

We conducted a complementary robustness check to assess whether LLMs make consistent judgments when directly reading full texts, compared to the extended summarization used in the paper for producing generalizable hypotheses under the limited context window. If using full texts from parsed PDFs, the results are essentially identical to those obtained with extended summaries (given their high fidelity), yielding a consistency score of 0.89 (i.e., the same judgment) across 1,000 judgments.

\section{Overall algorithmic framework}
\label{overall_al}

The overall algorithmic framework is shown in Algorithm \ref{algorithm}.

\begin{algorithm}[t]
\caption{\textbf{Hypothesis Search}}
\label{algorithm}
Initialize an empty hypothesis set $H$ and unexplained set $W$ (initially all pairs)\;
\For{each time step $t$}
{  
    Generate 5 hypotheses on top of the existing set $H$, from 50 random pairs of papers within the unexplained set $W$\;
    Update hypothesis set $H$\;
    Apply 5 new hypotheses to all pairs in $W$ by 3-fold confidence-weighted voting\;
    Apply 5 new hypotheses to explained pairs by 3-fold confidence-weighted voting as well\;
    Update the unexplained sample set $W$\;
    \If{$W < 5\%$ of all samples }
    {
        Stop searching\;
    }
}
Compare searched hypotheses with LLM priors extracted by self-talk\;
\end{algorithm}

\section{Related work}
\label{appendix:related_work}

\textbf{Hypothesis Generation}: Recent studies have explored the capabilities of LLMs in generating novel, testable, and interpretable hypotheses across a range of domains. Zhou, Liu et al. \citep{liu2024literature, liu2025hypobench, zhou2024hypothesis} proposed a framework that prompts LLMs to generate hypotheses from labeled datasets, beginning with a few examples. These initial hypotheses are iteratively refined using a reward mechanism inspired by the upper confidence bound algorithm from multi-armed bandits, balancing exploration and exploitation. Other works have integrated external knowledge — such as scientific literature, knowledge graphs, and raw web corpus — into the hypothesis generation process \citep{xiong2024improving, kulkarni2025scientific, wang2023learning, wang2024scimon, yang2023large}, improving the accuracy and novelty of LLM-generated hypotheses while reducing hallucinations. This line of work demonstrates the potential of LLMs in advancing scientific discovery. Beyond computer science, other disciplines are also beginning to explore these possibilities. Ludwig and Mullainathan \citep{ludwig2024machine} highlighted the promise of LLMs as tools for hypothesis generation in economics, arguing that machine learning models can uncover patterns and relationships that traditional econometric methods may miss. Manning, Zhu, and Horton \citep{manning2024automated} presented a simulation-based approach for automatically generating and testing social science hypotheses in silico. In each simulated agent-interaction case, causal relationships are both proposed and tested by the LLM system, with results aligning well with those derived from economic theory, particularly in predicting the signs of estimated effects. In medicine, LLMs have been used to hypothesize synergistic drug combinations for breast cancer treatment, with several predictions later validated in laboratory experiments \citep{bazgir2024agentichypothesis}, showcasing the potential of LLMs to surface effective, previously overlooked therapeutic strategies.

\textbf{Implicit Bias in LLMs}: Concerns about bias in AI systems have long focused on the ways these technologies can absorb and amplify harmful content from large-scale human-generated datasets -- particularly prejudices related to race, gender, and other social dimensions \citep{ntoutsi2020bias}. Large language models (LLMs), trained on vast corpora of human text, are especially prone to perpetuating and even intensifying these biases \citep{weidinger2021ethical, gallegos2024bias}. Such biases are not computational artifacts but reflect the deeper cultural and social frameworks that shape human communication \citep{babaeianjelodar2020quantifying, lee2024large, guilbeault2024online}. The recognition that LLMs encapsulate and propagate societal biases and heuristics is now well-established in literature, framing these models as inherently cultural and social technologies \citep{blodgett2020language, farrell2025large, weidinger2021ethical, naous2023having, li2024culturellm}. Much of the existing research on LLM bias has concentrated on explicit, easily identifiable forms of bias -- such as derogatory language, disparate system performance, erasure, exclusionary norms, misrepresentation, stereotyping, and toxicity \citep{gallegos2024bias, czarnowska2021quantifying}. Meanwhile, efforts in AI alignment, including outer alignment, inner alignment, and interpretability, have sought technological strategies for migrating such risks \citep{shen2023large, meade2021empirical, kumar2022language}. Despite progress in addressing explicit bias, unwritten codes in LLMs' "cognitive" process remain understudied. Like human cognition, LLMs may exhibit subtle, unconscious forms of bias even when explicit prejudices are absent \citep{shen2023large, lin2025implicit}. These implicit biases are often harder to detect yet can have deeper and more pervasive effects on model outputs and downstream decision-making. Thus, systematically identifying these ``unwritten codes" in LLMs is crucial to ensuring fairness, accuracy, and trustworthiness \citep{lin2025implicit}. One method worth mentioning in this topic is \citep{bai2025explicitly}. The authors draw on experimental paradigms from the Implicit Association Test (IAT) in psychology to design prompts that reveal latent biases in LLMs. Rather than directly asking models about their views on particular social groups -- which, like humans, they often avoid answering -- they instead present structured tasks that require associations or comparative decisions. For instance, a model might be asked to assign positive or negative words to names with racial or gender connotations, or to choose which of two candidates is more suitable for a specific role. This prompt-based approach bypasses explicit self-reports and mirrors the logic of involuntary responses in psychological testing, allowing underlying implicit stereotypes to surface indirectly through model behavior. 

\textbf{Science, Innovation, and LLM-as-a-Judge}: Prior work has increasingly applied LLMs to assist researchers by generating feedback, reviews, or reflections throughout the research process \citep{yuksekgonul2024textgrad, janssen2023use}. Most recent studies have focused on addressing challenges posed by the rapid growth of academic publishing, such as rising review workloads and declining review quality \citep{azad2024publication, rogers2023program}. Liang et al. \citep{liang2024can} evaluate GPT-4’s capacity to generate scientific feedback and find substantial overlap with human reviews, especially in early manuscript stages. Thakkar et al. \citep{thakkar2025llm} developed a "Review Feedback Agent" combining multiple LLMs; in a randomized trial, 27\% of reviewers revised their feedback after receiving LLM suggestions, which were more detailed and informative. Tyser et al. \citep{tyser2024ai} propose a broader framework, training LLMs to align with human preferences through pairwise comparisons and enhancing review robustness with techniques like adaptive prompting and role-playing. While prior studies on LLMs in peer review have primarily focused on augmenting or replicating human judgment to address scalability and quality concerns, our argument departs from this substitutional framing. Instead of merely simulating reviewers to reduce workload or enhance coverage of aspects, we leverage LLMs’ generative capacity to surface the implicit heuristics and tacit values that often shape peer review decisions but remain unspoken. By prompting LLMs to hypothesize why one paper is rated more highly than another, we reveal not only individual reviewer preferences but also broader cultural norms and evaluative codes that govern scientific judgment. In doing so, we treat AI not as a replacement for human thinking but as a diagnostic tool -- one that can reflect, exaggerate, and ultimately make visible the "unwritten code" underpinning peer review and, more broadly, other sociotechnical systems.

\section{Details of the dataset of conference submissions}
\label{appendix:ap2}
This dataset includes 26,731 submissions to 46 conferences: 1st ContinualAI Unconference, AAAI Conference on Artificial Intelligence 2024 and 2025, Symposium on Advances in Approximate Bayesian Inference 2024, Conference on Language Modeling 2024, Cooking Robotics Workshop 2024, Conference on Robot Learning 2023 and 2024,  Conference on Parsimony and Learning 2024, IEEE/CVF Conference on Computer Vision and Pattern Recognition 2023 and 2024, Workshop on Distributed Infrastructure for Common Good 2023, Workshop on Embodiment-Aware Robot Learning 2024, European Conference on Computer Vision 2024, Conference on Empirical Methods in Natural Language Processing 2023, European Space Power Conference 2023, Fast, Low-resource, and Accurate Organ and Pan-cancer Segmentation in Abdomen CT 2023, SIGIR Workshop on Generative Information Retrieval 2024, ACM/IEEE International Conference on Human-Robot Interaction 2023 and 2024, International Conference on Integration of Science and Technology for Sustainable Development 2024, International Conference on Learning Representations 2017 to 2025, International Conference on Machine Learning 2023 and 2024, ACM International Conference on the Theory of Information Retrieval 2024, International Joint Conference on Artificial Intelligence 2024, International Semantic Web Conference 2024, ACM SIGKDD Conference on Knowledge Discovery and Data Mining 2023 and 2024, ACM International Conference on Multimedia 2024, Conference on Neural Information Processing Systems 2021-2024, Neuro-Symbolic Learning and Reasoning in the era of Large Language Models 2024, Next-generation Data Governance Workshop 2024, Tsinghua University Advanced Machine Learning 2024.

Each paper in the dataset is represented with a structured format while retaining details. Each part in the representation is generated by feeding the corresponding sections into the LLM.

\begin{tcolorbox}[enhanced,
  sharp corners=south,
  colframe=black,
  colback=white,
  boxrule=1pt,
  arc=3mm,
  width=0.48\textwidth,
  title=\textbf{Case Study 2:},
  fonttitle=\bfseries,
  coltitle=black,
  breakable
]

\textbf{Title}:

Learnability of convolutional neural networks for infinite dimensional input via mixed and anisotropic smoothness
\end{tcolorbox}

\begin{tcolorbox}[enhanced,
  sharp corners=south,
  colframe=black,
  colback=white,
  boxrule=1pt,
  arc=3mm,
  width=0.48\textwidth,
  title=\textbf{Case Study 2:},
  fonttitle=\bfseries,
  coltitle=black,
  breakable
]
\textbf{Raw abstract}:

Among a wide range of success of deep learning, convolutional neural networks have been extensively utilized in several tasks such as speech recognition, image processing, and natural language processing, which require inputs with large dimensions. Several studies have investigated function estimation capability of deep learning, but most of them have assumed that the dimensionality of the input is much smaller than the sample size. However, for typical data in applications such as those handled by the convolutional neural networks described above, the dimensionality of inputs is relatively high or even infinite. In this paper, we investigate the approximation and estimation errors of the (dilated) convolutional neural networks when the input is infinite dimensional. Although the approximation and estimation errors of neural networks are affected by the curse of dimensionality in the existing analyses for typical function spaces such as the Holder and Besov spaces, we show that, by considering anisotropic smoothness, they can alleviate exponential dependency on the dimensionality but they only depend on the smoothness of the target functions. Our theoretical analysis supports the great practical success of convolutional networks. Furthermore, we show that the dilated convolution is advantageous when the smoothness of the target function has a sparse structure.

\end{tcolorbox}

\begin{tcolorbox}[enhanced,
  sharp corners=south,
  colframe=black,
  colback=white,
  boxrule=1pt,
  arc=3mm,
  width=0.48\textwidth,
  title=\textbf{Case Study 2:},
  fonttitle=\bfseries,
  coltitle=black,
  breakable
]

\textbf{Contextual background}:

Convolutional Neural Networks (CNNs) are widely used in applications like image recognition, speech processing, and natural language understanding. These applications often involve inputs that are extremely high-dimensional—so high, in fact, that they can be treated as infinite-dimensional in theory. However, most theoretical research on neural networks has focused on low- or fixed-dimensional settings, assuming the input has fewer dimensions than the number of samples available for training. Existing theoretical results on neural networks, such as those studying their function approximation capabilities, break down or become less meaningful in infinite-dimensional cases. Some prior studies attempted to address this by assuming inputs lie on a low-dimensional manifold or that the target function has certain regularity properties. Still, none fully addressed the behavior of CNNs when inputs are truly infinite-dimensional. This paper fills that gap. It explores whether CNNs can still learn effectively in such extreme settings by assuming the target function behaves differently across input dimensions, a property called anisotropic or mixed smoothness.

\end{tcolorbox}

\begin{tcolorbox}[enhanced,
  sharp corners=south,
  colframe=black,
  colback=white,
  boxrule=1pt,
  arc=3mm,
  width=0.48\textwidth,
  title=\textbf{Case Study 2:},
  fonttitle=\bfseries,
  coltitle=black,
  breakable
]

\textbf{Key idea}:

The main idea of the paper is that CNNs, especially those using dilated convolutions, can effectively learn even when the input is infinite-dimensional, as long as the target function has a particular kind of smoothness across dimensions. Rather than relying on the number of input dimensions, the authors argue that how the target function changes across different input coordinates is what truly determines learnability. This is formalized using two smoothness conditions: Mixed smoothness, where the function may vary differently across each input direction but still follows a structured form. Anisotropic smoothness, where the function is especially smooth (or insensitive) in certain input directions and more complex in others. The paper introduces a formal function space to describe such functions and then proves that CNNs can achieve good learning performance in these spaces, without being affected by the sheer dimensionality of the input. The key is that the CNNs, particularly with dilated convolutions, are capable of automatically detecting and focusing on the most relevant input components.

\end{tcolorbox}

\begin{tcolorbox}[enhanced,
  sharp corners=south,
  colframe=black,
  colback=white,
  boxrule=1pt,
  arc=3mm,
  width=0.48\textwidth,
  title=\textbf{Case Study 2:},
  fonttitle=\bfseries,
  coltitle=black,
  breakable
]

\textbf{Theoretical and methodological details}:

This section contains the core technical contributions of the paper. A. Learning Problem The paper studies a nonparametric regression problem, where the goal is to learn a function that maps an infinite-dimensional input to a real-valued output. The inputs can be thought of as sequences or functions, such as raw waveforms, high-resolution images, or long text embeddings. The challenge is to characterize how well a neural network can approximate this unknown function and how much error we can expect when training from a finite number of samples. B. Smoothness-Based Function Space To handle infinite-dimensional inputs, the authors define a new function space where functions are categorized based on their smoothness. Crucially, this smoothness is allowed to vary across input coordinates: Some directions may be very "rough" (i.e., important to capture precisely), while others can be ignored or treated coarsely. This idea mirrors how images contain both low-frequency (broad structure) and high-frequency (detail) components, and not all are equally important. By introducing a flexible way to quantify how much a function varies in each input direction, the authors build a mathematical foundation for analyzing the performance of neural networks in these spaces. C. Approximation Capabilities The paper first analyzes fully connected neural networks and shows that they can approximate functions in the proposed smoothness space well—but only if the network is explicitly given the most important input dimensions. In other words, these networks are theoretically powerful but need help in selecting relevant inputs. In contrast, CNNs can discover the important directions automatically thanks to their structured and hierarchical nature. The authors show that: Ordinary CNNs perform well when the smoothness of the target function increases gradually across input directions (e.g., the first few coordinates are rough, the rest smooth). Dilated CNNs are even more powerful when the target function has a sparse structure—that is, only a few input directions matter significantly, and these may be far apart. The authors prove that in both settings, CNNs can achieve fast convergence rates—meaning the prediction error decreases quickly as more data is observed—even if the input is infinite-dimensional. These rates depend only on the smoothness of the function, not on the number of input dimensions. D. Estimation from Data The paper also analyzes how well neural networks trained on data (using empirical risk minimization) can learn the true function. They show that: CNNs achieve dimension-independent learning rates under the right smoothness assumptions. For functions with gradually increasing smoothness, shallow CNNs suffice. For sparsely smooth functions, deeper or dilated CNNs are needed to identify and capture distant important features. This theoretical result aligns well with practical intuitions about CNNs being good at capturing both local and global structures.

\end{tcolorbox}

\begin{tcolorbox}[enhanced,
  sharp corners=south,
  colframe=black,
  colback=white,
  boxrule=1pt,
  arc=3mm,
  width=0.48\textwidth,
  title=\textbf{Case Study 2:},
  fonttitle=\bfseries,
  coltitle=black,
  breakable
]

\textbf{Experiments and results}:

The paper does not include any empirical experiments—no training on real datasets, no benchmarks, no runtime results. Instead, it provides a theoretically rigorous analytical framework to evaluate model performance, focusing on: Approximation error: How well the neural network class can represent the true function. Estimation error: How close the trained model gets to the true function when trained on finite data. The authors derive explicit mathematical expressions for the convergence rates of neural networks. These rates are shown to be: Polynomial in the number of samples, not exponential in the input dimension. Optimal or near-optimal, matching or extending existing bounds in lower-dimensional settings. They also compare their rates with those from other methods like kernel regression or traditional non-deep learning techniques, demonstrating that CNNs offer better or comparable rates in infinite-dimensional settings—especially when function smoothness is sparse or non-uniform. This analytical evaluation shows that CNNs, especially those using dilated convolutions, are theoretically justified for high- and infinite-dimensional data, even in the absence of experimental validation.

\end{tcolorbox}

\begin{tcolorbox}[enhanced,
  sharp corners=south,
  colframe=black,
  colback=white,
  boxrule=1pt,
  arc=3mm,
  width=0.48\textwidth,
  title=\textbf{Case Study 2:},
  fonttitle=\bfseries,
  coltitle=black,
  breakable
]

\textbf{Mentioned impact}:

This paper makes a theoretical contribution by offering comprehensive analyses of deep learning's learnability in infinite-dimensional spaces. Its results: Support the practical success of CNNs in real-world tasks involving high-dimensional data. Justify the use of dilated convolutions as an effective tool for handling sparse or long-range dependencies. Advance our understanding of when and why deep learning avoids the curse of dimensionality, shifting the focus from input size to function structure.

\end{tcolorbox}

\section{Prompts used in the paper}
\label{appendix:prompts}

\begin{tcolorbox}[enhanced,
  sharp corners=south,
  colframe=black,
  colback=lightgray,
  boxrule=1pt,
  arc=3mm,
  width=0.48\textwidth,
  title=\textbf{Case Study 2:},
  fonttitle=\bfseries,
  coltitle=black,
]
\#Hypothesis evaluation\\
You are a computer scientist.\\
Task: Evaluate the following hypothesis based on the content of two papers.\\
Hypothesis: "\{hypothesis\}"\\
Paper 1 Content: \{row['comprehensive content 1']\}\\
Paper 2 Content: \{row['comprehensive content 2']\}\\
Instructions:\\
1. Analyze the content from both papers carefully to determine whether Paper 2 is superior in the aspects identified in the hypothesis.\\
2. Provide your response in the exact JSON format (without additional commentary):\\\\
\texttt{<label>X</label>}\\\\[1em]
\texttt{<confidence>Y</confidence>}\\\\[1em]
- X: "1" if Paper 2 is superior in the identified aspects (i.e., the comparison of contents supports the hypothesis), or "0" if not.\\
- Y: An integer between 0 and 10 indicating your confidence level in this judgment (with 10 meaning highest certainty).
\end{tcolorbox}

\begin{tcolorbox}[enhanced,
  sharp corners=south,
  colframe=black,
  colback=lightgray,
  boxrule=1pt,
  arc=3mm,
  width=0.48\textwidth,
  title=\textbf{Case Study 2:},
  fonttitle=\bfseries,
  coltitle=black,
]
\#Hypothesis iterative search\\
You are a computer scientist.\\
You are analyzing two papers submitted to the same venue.\\
Paper 1 (comprehensive\_content\_1) received a lower peer review score (rating\_1) than Paper 2 (comprehensive\_content\_2 and rating\_2).\\
Your task is to generate 5 specific, creative, and critical hypotheses explaining why Paper 1 may have received a lower score.\\\\
- Base your hypotheses on a rigorous content-driven comparison of the provided dataset {df}, identifying concrete differences that might have influenced peer review scores.\\
- Avoid vague reasoning and ensure each hypothesis is distinct, well-reasoned, and actionable.\\
- Ensure that the generated hypotheses are distinct from those in {df\_hypothesis}.\\

Provide your response in the exact JSON format (without additional commentary):\\
"Compared to another paper, one paper [your hypothesis]."
\end{tcolorbox}

\begin{tcolorbox}[enhanced,
  sharp corners=south,
  colframe=black,
  colback=lightgray,
  boxrule=1pt,
  arc=3mm,
  width=0.48\textwidth,
  title=\textbf{Hypothesis Generation Prompt},
  fonttitle=\bfseries,
  coltitle=black,
]
\#Prior hypothesis search\\
You are a computer scientist.\\
Task: Generate specific, creative, and critical hypotheses about concrete differences between two papers that might have influenced their peer review scores.
You will perform 4 rounds. In each round, generate 5 hypotheses.\\[1em]

Each hypothesis must:

- Focus on a distinct, well-defined, and actionable factor.

- Avoid vague reasoning.\\

Provide your response in the exact JSON format (without additional commentary):\\
"Compared to another paper, one paper [your hypothesis]."


\end{tcolorbox}

\begin{tcolorbox}[enhanced,
  sharp corners=south,
  colframe=black,
  colback=lightgray,
  boxrule=1pt,
  arc=3mm,
  width=0.48\textwidth,
  title=\textbf{Feedback Annotation Task},
  fonttitle=\bfseries,
  coltitle=black,
]
\#Annotation of human review comments\\
\#attach examples provided to LLMs for three conditions across 20 hypotheses\\
You are a computer scientist.\\
Please annotate the human feedback below by mapping it to the aspects in the 20 hypotheses.\\
Return your answer in exactly the following JSON format (no extra text or commentary):\\[0.5em]
\texttt{\{"scores": [X1, X2, X3, ..., X20]\}}\\[0.5em]
Each Xi must be one of: \texttt{-1}, \texttt{0}, or \texttt{1}\\
- Use \texttt{1} if the feedback praises the standard\\
- Use \texttt{-1} if the feedback criticizes the standard\\
- Use \texttt{0} if the standard is not mentioned\\[1em]

\textbf{Hypotheses:}\\
\texttt{{\{hypothesis\_list\}}}\\[1em]

\textbf{Feedback:}\\
\texttt{{\{feedback\_text\}}}
\end{tcolorbox}

\section{The list of generated hypotheses}
The list of generated hypotheses is shown in Table \ref{table:hypos}.

 \section{Cross-validation dictionary}
\label{appendix:dict}

The dictionary method used in the paper to complement LLM labeling/matching is shown in Table \ref{table:dictionary_table}. Note that this heuristic dictionary-like approach, while widely used in social science \citep{guo2016big, frankel2022disclosure}, ultimately serves as a substitute rather than the major methods of labeling human comments of "mention" or matching priors and posteriors, presented in the paper.

 \section{Humans explicitly reward the internal qualities of science, but a significant hidden driver remains.}
\label{appendix:regression}

We use GPT-4o with few-shot to annotate human review comments across the aspects of 20 hypotheses. Each aspect is labeled as praise (1), not mentioned (0), or criticism (-1). We then run an OLS regression (using \texttt{scipy.stats} package) to examine how reviewers’ attitudes toward these aspects relate to the scores they assign within the conference. The results are shown in Table \ref{regression}.

We have two points to discuss:

\begin{itemize}
\item Only five hypotheses are statistically significant meaningful predictors of human review scores, and all have positive coefficients—indicating that when these aspects mentioned in the hypothesis are positive, the associated paper tends to receive a higher score. These hypotheses (highlighted in red) clearly reflect the internal qualities of scientific work. This suggests that reviewers do use their scores to explicitly reward normative aspects of science. Notably, two of these five hypotheses—theoretical rigor and the balance between theoretical analysis and practical impact—also appear among the top five hypotheses favored by priors but not favored by posteriors.
\item However, the regression model's explanatory power is very low (\(R^2 = 0.06\)). This suggests that much of what drives reviewers' scoring decisions is not captured by the explicit reward of these normative qualities. There are implicit factors strongly influencing the review process that are not reflected directly in the scores.
\end{itemize}

 \section{Discussions of position bias in LLM-as-a-judge}
\label{appendix:position}

To address potential position bias, we adopt two mitigation strategies in the original paper. First, we construct paper pairs with substantial quality differences, as prior research has shown that position bias is most pronounced when the quality gap is small \citep{li2024split, shi2024judging}. Second, we randomize the positions of Paper 1 and Paper 2 across voting, while ensuring logically consistent prompts and vote aggregation. We then assess whether position bias is indeed stronger when the quality signal gap is small. Specifically, we regress the consistency rate — defined as the proportion of consistent judgments across 20 hypotheses after swapping paper positions — on the review score gap between the two papers. Our analysis (Table \ref{position_table}) shows that a larger review score gap is indeed associated with greater consistency, and the effect is statistically significant, though weak in magnitude.

\begin{table*}[ht!]
    \resizebox{\textwidth}{!}{
    \renewcommand{\arraystretch}{1} 
    \setlength{\tabcolsep}{6pt}        
    \centering                      
    \begin{tabular}{|p{6cm}|c|c|c|p{2.3cm}|p{2.3cm}|}   
        \hline
        \textbf{Hypothesis} & \textbf{Round} & \textbf{Prior} & \textbf{Posterior} & \textbf{Rank in shifts} & \textbf{Mention} \\
        \hline
        lack justification regarding its novelty & 1 & 0.36 & 0.20 & 13 & 0.67 \\
         \hline
        \textbf{\textcolor{blue}{not theoretically rigorous}} & 1 & 0.92 & 0.18 & 20 & 0.57 \\
         \hline
        \textbf{\textcolor{blue}{the design is over-engineered, not elegant, and unnecessarily complicated}} & 1 & 0.50 & 0.02 & 18 & 0.74 \\
         \hline
        not use a fair benchmark for evaluation & 1 & 0.32 & 0.08 & 15 & 0.16 \\
         \hline
        \textbf{\textcolor{blue}{lack implementation and reproducibility details}} & 1 & 0.44 & 0.18 & 16 & 0.05 \\
        \hline
        lack comprehensive ablation and hyperparameter analysis & 2 & 0.58 & 0.35 & 14 & 0.26 \\
         \hline
         the contribution is incremental & 2 & 0.06 & 0.24 & 9 & 0.12 \\
         \hline
        \textbf{\textcolor{blue}{ not balance theoretical justification and practical impact}} & 2 & 0.68 & 0.20 & 17 & 0.60 \\
         \hline
        \textbf{\textcolor{orange}{the presentation is too complicated}} & 2 & 0.14 & 0.50 & 4 & 0.45 \\
         \hline
        \textbf{\textcolor{orange}{unclear about its contextualization within the related literature}} & 2 & 0.30 & 0.65 & 5 & 0.50 \\
        \hline
         misaligned with the emerging scope & 3 & 0.14 & 0.10 & 11 & 0.11 \\
         \hline
        lack empirical evidence in terms of model and architecture design choice & 3 & 0.16 & 0.51 & 6 & 0.18 \\
         \hline
        limited in scalability to real-world scenarios & 3 & 0.44 & 0.35 & 12 & 0.20 \\
         \hline
        \textbf{\textcolor{blue}{insufficient in testing robustness such as uncertainty and bias in evaluation experiments}} & 3 & 0.84 & 0.28 & 19 & 0.53 \\
         \hline
        \textbf{\textcolor{orange}{insufficiently discuss social and ethical aspects of its algorithms}} & 3 & 0.18 & 0.55 & 3 & 0.02 \\
        \hline
        weakly motivated in its training and optimization techniques & 4 & 0.30 & 0.62 & 7 & 0.82 \\
         \hline
        the theoretical assumptions are unrealistic & 4 & 0.40 & 0.58 & 8 & 0.61 \\
         \hline
        poorly explore edge cases and failure modes & 4 & 0.58 & 0.60 & 10 & 0.24 \\
         \hline
       \textbf{\textcolor{orange}{require dense prior background knowledge}} & 4 & 0.02 & 0.59 & 2 & 0.07 \\
         \hline
       \textbf{\textcolor{orange}{not interact well with adjacent domains, such as statistics and cognitive science}} & 4 & 0.06 & 0.77 & 1 & 0.01 \\
        \hline
    \end{tabular}
    }
    \caption{The generated hypotheses. A shift in rank indicates the change in "importance", measured as posterior minus prior (i.e., a high shift means the hypotheses that were not highly weighted in the prior but were used much more frequently in actual judgments). Orange highlights the top 5 hypotheses with the largest gains (e.g., contextual and narrative perspectives), while blue highlights the top 5 with the largest losses (e.g., theoretical rigor and normative perspectives). "Mention" indicates the proportion of human comments that mentioned the hypothesis, as identified by GPT-4o. All hypotheses start with "Compared to the other, one paper..."}
    \label{table:hypos}
\end{table*}

\newpage

\begin{table*}[ht]
 \renewcommand{\arraystretch}{1} 
 \setlength{\tabcolsep}{6pt}        
 \centering                      
  \begin{tabular}{|c|p{12cm}|}   
  \hline
 \textbf{Hypothesis Index} & \textbf{Dictionary} \\
 \hline
 1 & ["novelty", "novel"] \\ 
 \hline
  2 & ["rigor", "rigorous", "theoretically rigorous", "theoretical rigor", "proof"] \\
 \hline
  3 & ["over-engineering", "over-engineered", "elegance", "elegant", "complicated model", "complicated architecture", "clarity"] \\
  \hline
  4 & ["fair benchmark", "benchmarking", "evaluation"] \\
  \hline
  5 & ["reproducibility", "implementation detail"] \\
 \hline
  6 & ["ablation", "hyperparameter"] \\
  \hline
  7 & ["incremental", "A+B"] \\
  \hline
  8 & ['theoretical justification', "practical impact"] \\
  \hline
 9 & ["presentation", "readability", "writing"] \\
  \hline
 10 & ["related work", "contextualization", "literature"] \\
 \hline
 11 & ["emerging", "out of date", "scope", "trend"] \\
  \hline
  12 & ["design choice", "empirical evidence", "architecture design", "model design"] \\
  \hline
  13 & ["scalability", "real-world", "applicability"] \\
 \hline
  14 & ["uncertainty", "biased evaluation", "evaluation bias", "evaluation", "robustness"] \\
  \hline
  15 & ["societal", "ethical", "society", "ethic", "social"] \\
  \hline
  16 & ["optimization", "training", "train", "loss function"] \\
 \hline
 17 & ["theoretical assumption", "practical", "practically", "in practice", "unrealistic assumption"] \\
 \hline
 18 & ["failure mode", "edge case"] \\
 \hline
 19 & ["implicit knowledge", "prior knowledge", "implicit background", "prior background", "background knowledge"] \\
 \hline
 20 & ["interdisciplinary", "multidisciplinary", "statistics", "cognitive science", "neural science", "psychology", "cognition"] \\
 \hline
 \end{tabular}
    \caption{The dictionary method used in the paper to complement LLM labeling/matching.}
    \label{table:dictionary_table}
\end{table*}

\newpage

\begin{table*}[ht!]
\centering
\caption{Regression results on the mentioned perspectives and human scoring.}
\label{regression}
\resizebox{1\textwidth}{!}{
\begin{tabular}{p{6cm}|rrrrr}
\toprule
Hypothesis & Coef. & Std. Err. & t & P$>$$|t|$ & [0.025, 0.975] \\
\midrule
const            & 5.966  & 0.100 & 59.508 & 0.000 & [5.769, 6.163] \\
 \hline
\hl{\textbf{lack justification regarding its novelty}}  & 0.231  & 0.096 &  2.410 & 0.016 & [0.043, 0.420] \\
 \hline
\hl{\textbf{not theoretically rigorous}}  & 0.209  & 0.095 &  2.194 & 0.028 & [0.022, 0.396] \\
 \hline
 the design is over-engineered, not elegant, and unnecessarily complicated   & 0.046  & 0.100 &  0.463 & 0.643 & [-0.150, 0.242] \\
  \hline
lack comprehensive ablation and hyperparameter analysis  & 0.152  & 0.087 &  1.754 & 0.080 & [-0.018, 0.322] \\
 \hline
unclear about its contextualization within the related literature  & 0.069  & 0.082 &  0.836 & 0.403 & [-0.093, 0.230] \\
 \hline
\hl{\textbf{the contribution is incremental}} & 0.205  & 0.089 &  2.310 & 0.021 & [0.031, 0.379] \\
 \hline
\hl{\textbf{not use a fair benchmark for evaluation}}  & 0.169  & 0.085 &  1.998 & 0.046 & [0.003, 0.336] \\
 \hline
the presentation is too complicated  & -0.006 & 0.081 & -0.078 & 0.938 & [-0.166, 0.153] \\
 \hline
lack implementation and reproducibility details & 0.019  & 0.098 &  0.198 & 0.843 & [-0.172, 0.211] \\
 \hline
\hl{\textbf{not balance theoretical justification and practical impact}} & 0.356  & 0.136 &  2.619 & 0.009 & [0.089, 0.622] \\
 \hline
misaligned with the emerging scope & 0.193  & 0.223 &  0.865 & 0.388 & [-0.245, 0.630] \\
 \hline
 insufficiently discuss social and ethical aspects of its algorithms & -0.051 & 0.262 & -0.196 & 0.844 & [-0.565, 0.462] \\
  \hline
not interact well with adjacent domains, such as statistics and cognitive science  & -0.081 & 0.249 & -0.325 & 0.745 & [-0.569, 0.408] \\
 \hline
lack empirical evidence in terms of model and architecture design choice  & 0.015  & 0.102 &  0.149 & 0.882 & [-0.185, 0.215] \\
 \hline
limited in scalability to real-world scenarios & 0.034  & 0.110 &  0.308 & 0.758 & [-0.183, 0.251] \\
 \hline
insufficient in testing robustness such as uncertainty and bias in evaluation & 0.026  & 0.204 &  0.127 & 0.899 & [-0.374, 0.426] \\
 \hline
weakly motivated in its training and optimization techniques & 0.073  & 0.169 &  0.430 & 0.668 & [-0.259, 0.404] \\
 \hline
the theoretical assumptions are unrealistic & 0.041  & 0.175 &  0.232 & 0.817 & [-0.302, 0.383] \\
 \hline
poorly explore edge cases and failure modes & -0.030 & 0.080 & -0.374 & 0.708 & [-0.186, 0.127] \\
  \hline
require dense prior background knowledge & 0.006  & 0.099 &  0.056 & 0.955 & [-0.189, 0.200] \\
\bottomrule
\end{tabular}
}
\end{table*}

\newpage

\begin{table}[htbp]
\centering
\caption{OLS regression results on the position bias. Standard errors assume that the covariance matrix of the errors is correctly specified.}
\label{position_table}
\begin{tabular*}{\columnwidth}{lclc}
\toprule
\textbf{Dep. Variable:}    &   consistency   & \textbf{R-squared:}      & 0.015 \\
\textbf{Model:}            &       OLS       & \textbf{Adj. R-squared:} & 0.015 \\
\textbf{Method:}           &  Least Squares  & \textbf{F-statistic:}    & 77.26 \\
\textbf{Prob (F-statistic):} & 2.03e-18      & \textbf{Log-Likelihood:} & 5245.2 \\
\textbf{No. Observations:} & 5000            & \textbf{AIC:}            & -1.049e+04 \\
\textbf{Df Residuals:}     & 4998            & \textbf{BIC:}            & -1.047e+04 \\
\textbf{Df Model:}         & 1               &                          &          \\
\textbf{Covariance Type:}  & nonrobust       &                          &          \\
\bottomrule
\end{tabular*}

\vspace{0.5em}

\begin{tabular*}{\columnwidth}{lcccccc}
               & \textbf{coef} & \textbf{std err} & \textbf{t} & \textbf{P$>|$t$|$} & \textbf{[0.025} & \textbf{0.975]} \\
\midrule
\textbf{const} & 0.74 & 0.00 & 237.00 & 0.00 & 0.73 & 0.75 \\
\textbf{gap}   & 0.01 & 0.00 & 8.79   & 0.00 & 0.01 & 0.01 \\
\bottomrule
\end{tabular*}

\vspace{0.5em}

\begin{tabular*}{\columnwidth}{lclc}
\textbf{Omnibus:}       & 18.990 & \textbf{Durbin-Watson:} & 1.985 \\
\textbf{Prob(Omnibus):} & 0.000  & \textbf{Jarque-Bera (JB):} & 16.480 \\
\textbf{Skew:}          & -0.085 & \textbf{Prob(JB):}      & 0.000264 \\
\textbf{Kurtosis:}      & 2.777  & \textbf{Cond. No.:}     & 9.08 \\
\bottomrule
\end{tabular*}
\end{table}

\end{document}